\documentclass[preprint,12pt]{elsarticle}  
\usepackage{graphicx} 
\usepackage{amssymb} 
\usepackage{amsmath} 

\journal{Physics Letters B} 

\begin{document} 

\begin{frontmatter} 

\title{$\eta$-nuclear bound states revisited} 

\author[a]{E.~Friedman} 
\author[a]{A.~Gal\corref{cor1}} 
\cortext[cor1]{Corresponding author: Avraham Gal, avragal@vms.huji.ac.il} 
\author[b]{J.~Mare\v{s}} 
\address[a]{Racah Institute of Physics, The Hebrew University, 91904 
Jerusalem, Israel} 
\address[b]{Nuclear Physics Institute, 25068 \v{R}e\v{z}, Czech Republic} 

\begin{abstract} 
The strong energy dependence of the $s$-wave $\eta N$ scattering amplitude 
at and below threshold, as evident in coupled-channels $K$-matrix fits 
and chiral models that incorporate the $S_{11}$ $N^{\ast}(1535)$ resonance, 
is included self-consistently in $\eta$-nuclear bound-state calculations. 
This approach, applied recently in calculations of kaonic atoms and 
$\bar K$-nuclear bound states, is found to impose stronger constraints 
than ever on the onset of $\eta$-nuclear binding, with a minimum value of 
Re\;$a_{\eta N}\approx 0.9$~fm required to accommodate an $\eta$-$^{4}$He 
bound state. Binding energies and widths of $\eta$-nuclear states are 
calculated within several underlying $\eta N$ models for nuclei across the 
periodic table, including $_{~\eta}^{25}$Mg for which some evidence was 
proposed in a recent COSY experiment. 
\end{abstract} 


\end{frontmatter}

\section{Introduction} 
\label{intro} 

\begin{figure}[thb] 
\begin{center} 
%\rotatebox{270}{\includegraphics[height=9cm]{wycech.eps}} 
\includegraphics[width=0.6\textwidth]{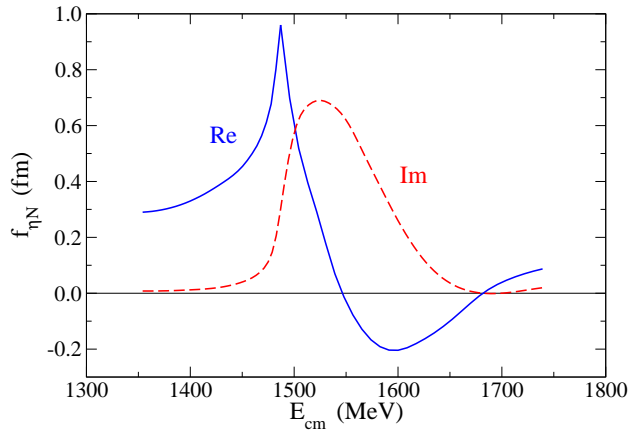} 
\caption{Energy dependence of the $s$-wave $\eta N$ cm scattering 
amplitude $f_{\eta N}(\sqrt{s})$, with scattering length 
$a_{\eta N}\equiv f_{\eta N}(\sqrt{s_{\rm th}})=0.91(6)+{\rm i}0.27(2)$~fm, 
as calculated by Green and Wycech (GW) fitting pion reaction and 
photoproduction data over a wide range of cm energies $E_{\rm cm}$ 
(=$\sqrt{s}$). The spike of Re\;$f$ is located at the $\eta N$ threshold, 
whereas Im\;$f$ peaks close to the $N^{\ast}(1535)$ mass. Figure adapted 
from Ref.~\cite{GW05}.} 
\label{fig:wycech} 
\end{center} 
\end{figure} 

Searches for meson-nuclear bound states have focused on $K^-$ and $\eta$ 
mesons, motivated by a general theoretical consensus that the near-threshold 
$\bar K N$ and $\eta N$ attraction generated by the $s$-wave resonances 
$\Lambda(1405)$ and $N^{\ast}(1535)$, respectively, translates into 
sufficiently attractive $K^-$-nucleus and $\eta$-nucleus interactions. 
A corollary of this resonance dominance is a strong energy dependence 
of the underlying $\bar K N$ and $\eta N$ interactions. Here we apply the 
lessons gained by handling the strong energy dependence of the near-threshold 
$\bar K N$ interaction in $K^-$-nuclear calculations \cite{Gal13} to 
$\eta$-nuclear bound-state calculations. Early calculations by Haider and 
Liu \cite{HL86,LH86} predicted $\eta$-nuclear bound states beginning with 
nuclear mass number $A\sim 12$. In these, as well as in a follow-up 
calculation \cite{HL02}, a fairly weak $\eta N$ attraction input was used, 
with Re\;$a_{\eta N}\lesssim 0.3$~fm, where $a_{\eta N}$ is the $\eta N$ 
scattering length. Several versions of coupled-channels chiral models 
\cite{weise95,IOV02,MBM12} give similar values as well, whereas other models, 
particularly those using $K$-matrix methods to fit $\pi N$ and $\gamma N$ 
reaction data in the $N^{\ast}(1535)$ resonance region, 
e.g. \cite{GW05,Arndt05,SLM13}, yield considerably stronger $\eta N$ 
attraction with values of Re\;$a_{\eta N}\approx 1$~fm.{\footnote{As for 
Im\;$a_{\eta N}$, it is constrained by $\pi N\to\eta N$ cross-section 
measurements, with values of Im\;$a_{\eta N}\sim 0.2$--0.3~fm in most 
theoretical analyses.}} This might suggest that the onset of $\eta$-nuclear 
binding occurs already in the He isotopes for which strong final-state 
interaction precursors have been noted in proton- and deuteron-initiated 
$\eta$ production \cite{LNS88,LNS94}. A robust pattern of $\eta$-nuclear 
bound states could yield useful information on the size of SU(3) flavor 
$\eta-\eta'$ mixing and about axial U(1) dynamics \cite{BTh06}. To date, 
however, experimental searches for such bound states have been unsuccessful, 
e.g. the latest negative results for $_{\eta}^{3}$He (in photoproduction 
on $^{3}$He \cite{MAMI12}) and for $_{\eta}^{4}$He (in $dd\to {^{3}{\rm He}}\;
p\pi^{-}$ \cite{COSY13}). 

Regardless of the strong model dependence of Re\;$a_{\eta N}$, all studies 
of the $\eta N$ system near threshold, $\sqrt{s_{\rm th}}=m_N+m_{\eta}
\approx 1487$ MeV, agree that both real and imaginary parts of the $s$-wave 
center-of-mass (cm) scattering amplitude $f_{\eta N}$ decrease steeply 
in going subthreshold, as illustrated in Fig.~\ref{fig:wycech}. 
Since the in-medium $\eta N$ interaction relevant to the evaluation of 
$\eta$-nuclear bound states involves subthreshold $\eta N$ configurations, 
a procedure for going subthreshold is mandatory. Previous calculations 
focused on shifting the energy variable of $f_{\eta N}(\sqrt{s})$ or its 
in-medium version by a fixed amount below threshold: $\delta \sqrt{s}=-30$~MeV 
was found in Ref.~\cite{HL02} to provide a good approximation to a variety of 
off-shell effects, whereas $\delta \sqrt{s}=-B_{\eta}$, with $B_{\eta}$ the 
$\eta$-nuclear binding energy, was used in Refs.~\cite{GR02,jido}. The latter 
procedure requires a self-consistent calculation to ensure that the $B_{\eta}$ 
argument of the input $f_{\eta N}$ coincides with the $B_{\eta}$ output of the 
binding energy calculation. However, it was shown in our recent studies of 
$K^-$-nuclear dynamics \cite{CFGGM11a,CFGGM11b,FG12,GM12,BGL12,FG13} that a 
more involved self-consistent calculation is required to correctly implement 
the subthreshold energy dependence, and it is this self-consistent procedure 
that is applied here to calculate $\eta$-nuclear bound states. This procedure 
results in imposing stronger constraints than ever on the onset of 
$\eta$-nuclear binding. 

\begin{table}[hbt] 
\begin{center} 
\caption{$\eta N$ scattering length $a_{\eta N}$ (in fm) in three 
coupled-channels models used in the present work. M1 and M2 correspond 
to versions I and II, respectively, of $\eta N$ amplitudes from the 
recent chiral-model work by Mai et al. \cite{MBM12}, and GW denotes the 
$K$-matrix $\eta N$ amplitude due to Green and Wycech \cite{GW05} shown in 
Fig.~\ref{fig:wycech}.} 
\begin{tabular}{lcc} 
%\hline 
\noalign{\smallskip}\hline\noalign{\smallskip} 
%\multicolumn{2}{c}{M1} & \multicolumn{2}{c}{M2} & \multicolumn{2}{c}{GW} \\
Model & Re\;$a_{\eta N}$ & Im\;$a_{\eta N}$ \\ 
%\hline 
\noalign{\smallskip}\hline\noalign{\smallskip} 
M1 & 0.22 & 0.24 \\ 
M2 & 0.38 & 0.20 \\ 
GW & 0.96 & 0.26 \\ 
\hline 
%\noalign{\smallskip}\hline 
\end{tabular} 
\label{tab:input} 
\end{center} 
\end{table} 

Below we proceed to describe briefly the 
self-consistent procedure used to handle the subthreshold energy dependence 
of the $\eta N$ amplitude for bound nucleons, and its embedding into 
a dynamical Relativistic Mean Field (RMF) scheme which allows for 
the first time to consider the polarization of the core nucleus by the bound 
$\eta$ meson. To span a broad range of bound-state scenarios we apply our 
methodology to three distinct $\eta N$ amplitude models, with threshold 
values listed in Table~\ref{tab:input}. These amplitudes differ primarily 
in the value of the real part, while their shape below threshold exhibits 
a substantial decrease particularly for Im\;$a_{\eta N}$, as illustrated 
for the GW amplitude model \cite{GW05} in Fig.~\ref{fig:wycech}. We have 
calculated $\eta$-nuclear bound states across the periodic table for these 
three amplitude models, as reported and discussed here for $1s_{\eta}$ states. 
Finally, we also confront our results with a recent experimental suggestion 
of a $_{~\eta}^{25}$Mg bound state \cite{Bud09}.

\section{Methodology} 
\label{sec:meth} 

In close analogy to the latest calculation of $K^-$-nuclear bound states 
\cite{GM12}, we calculate $\eta$-nuclear bound states by solving 
self-consistently the Klein-Gordon (KG) equation 
\begin{equation} 
[\:\nabla^2 + {\tilde\omega}_{\eta}^2 - m_{\eta}^2 - 
\Pi_{\eta}(\omega_{\eta},\rho)\:]\:\psi=0, 
\label{eq:KG} 
\end{equation} 
where ${\tilde\omega}_{\eta}=\omega_{\eta}-{\rm i}\Gamma_{\eta}/2$ and 
$\omega_{\eta}=m_{\eta}-B_{\eta}$, with $B_{\eta}$ and $\Gamma_{\eta}$ 
the binding energy and the width of the $\eta$-nuclear bound state. 
The self-energy operator $\Pi_{\eta}$ is related to a density- and 
energy-dependent optical potential $V_{\eta}$ which is given by the 
following ``$t\rho$" form:  
\begin{equation} 
\Pi_{\eta}(\omega_{\eta},\rho)\equiv 2\omega_{\eta}V_{\eta}=-4\pi 
%\frac{\sqrt{s}}{m_N} 
F_{\eta N}(\sqrt{s},\rho)\rho, 
\label{eq:Pi} 
\end{equation} 
where $s=(E_{\eta}+E_N)^2-({\vec p}_{\eta}+{\vec p}_N)^2$ is the Lorentz 
invariant Mandelstam variable $s$ which reduces to the square of the total 
$\eta N$ energy in the two-body cm frame and $F_{\eta N}$ is the in-medium 
$\eta N$ $s$-wave scattering amplitude in the lab system. Note that for 
$A \gg 1$ the lab system approximates well the $\eta$-nucleus cm system. 
Our in-medium $F_{\eta N}$ accounts for Pauli correlations in the 
Ericson-Ericson multiple-scattering approach, as reformulated in 
Ref.~\cite{WRW97} and used recently in Ref.~\cite{FG13}: 
\begin{equation} 
F_{\eta N}(\sqrt{s},\rho)=\frac{{\tilde f}_{\eta N}(\sqrt{s})}
{1+\xi(\rho){\tilde f}_{\eta N}(\sqrt{s})\rho}, \;\;\;\;\; 
\xi(\rho)=\frac{9\pi}{4p_F^2}, 
\label{eq:WRW} 
\end{equation} 
where ${\tilde f}_{\eta N}(\sqrt{s})=(\sqrt{s}/m_N)f_{\eta N}(\sqrt{s})$, 
with the kinematical factor $\sqrt{s}/m_N$ transforming $f$ from the 
two-body cm frame to the lab $\tilde f$, and where $p_F$ is the local 
Fermi momentum corresponding to density $\rho=2p_F^3/(3\pi^2)$. Note that 
$F_{\eta N}(\sqrt{s},\rho)\to {\tilde f}_{\eta N}(\sqrt{s})$ upon $\rho\to 0$, 
as required by the low-density limit. Extensions of Eq.~(\ref{eq:WRW}) 
to coupled channels and inclusion of self-energies do not change the 
results presented here in any qualitative way and will be discussed 
elsewhere \cite{CFGM13}. 

In specifying the two-body cm energy $\sqrt{s}$ appearing in 
Eq.~(\ref{eq:WRW}) we recall that 
$s=(\sqrt{s_{\rm th}}-B_{\eta}-B_N)^2-({\vec p}_{\eta}+{\vec p}_N)^2$, 
where the momentum-dependent term provides additional downward energy 
shift to that arising from the sum of binding energies $B_{\eta}+B_N$. 
Unlike in the free-space ${\eta}N$ cm system where 
$({\vec p}_{\eta}+{\vec p}_N)_{\rm cm}=0$, this term in the lab system 
was found to contribute substantially in realistic nuclear applications 
\cite{CFGGM11a,CFGGM11b}. It has been verified numerically by us that 
$({\vec p}_{\eta}+{\vec p}_N)^2$ is well approximated by its angle-average 
$(p_{\eta}^2+p_N^2)$. Near threshold, then, to leading order in binding 
energies and kinetic energies with respect to masses, one obtains 
\begin{equation} 
\sqrt{s} \approx \sqrt{s_{\rm th}} - B_N - B_{\eta} - \xi_N\frac{p_N^2}{2m_N} 
- \xi_{\eta}\frac{p_{\eta}^2}{2m_{\eta}}, 
\label{eq:approx} 
\end{equation} 
where $\xi_{N(\eta)}\equiv m_{N(\eta)}/(m_N+m_{\eta})$. To transform momentum 
dependence into density dependence, the nucleon kinetic energy $p_N^2/(2m_N)$ 
is approximated within the Fermi gas model by $T_N(\rho/\rho_0)^{2/3}$, with 
average bound-nucleon kinetic energy $T_N=23.0$ MeV, and the ${\eta}$ kinetic 
energy $p_{\eta}^2/(2m_{\eta})$ is substituted within the local density 
approximation by $-B_{\eta}-{\rm Re}\:V_{\eta}(\sqrt{s},\rho)$. 
Thus, the {\it in-medium} $\sqrt{s}=\sqrt{s_{\rm th}}+\delta\sqrt{s}$ energy 
argument of $F_{\eta N}$ in Eq.~(\ref{eq:WRW}) is density-dependent, with a 
form adjusted to respect the low-density limit, $\delta\sqrt{s}\to 0$ with 
$\rho\to 0$, as used recently in $K^-$-atom studies \cite{FG13}: 
\begin{equation} 
\delta\sqrt{s}\approx -B_N\frac{\rho}{{\bar\rho}}-
\xi_N B_{\eta}\frac{\rho}{\rho_0}-\xi_N T_N(\frac{\rho}{\rho_0})^{2/3}+
\xi_{\eta}{\rm Re}~V_{\eta}(\sqrt{s},\rho). 
\label{eq:sqrts} 
\end{equation} 
Here $B_N\approx 8.5$~MeV is an average nucleon binding energy and $\rho_0$ 
($\bar\rho$) is the maximal (average) nuclear density. The appearance of 
the $V_{\eta}$ term due to $p_{\eta}\neq 0$ in finite nuclei contrasts with 
the common assumption $p_{\eta}=0$ made in nuclear matter calculations. The 
dependence of $V_{\eta}$ on energy through $\sqrt{s}$ and on density $\rho$ is 
explicitly marked in this expression. Note that for an attractive $V_{\eta}$ 
and as long as $\rho\neq 0$, the shift of the two-body energy away from 
threshold is negative definite, $\delta\sqrt{s}<0$, even as $B_{\eta}\to 0$. 
For a given $B_{\eta}$, neither $\sqrt{s}$ nor $V_{\eta}$ can be 
evaluated separately, implying that $V_{\eta}$ is to be constructed 
{\it self-consistently} together with $\sqrt{s}$, which takes typically about 
5 cycles of iteration. Once $V_{\eta}(\sqrt{s},\rho)$ has been determined, 
it is used in the KG Eq.~(\ref{eq:KG}) to solve for the binding energy 
{\it eigenvalue} $B_{\eta}^{(n\ell)}$ in the $\eta$-nuclear $n\ell$ 
single-particle state. While varying the value of $B_{\eta}^{(n\ell)}$ 
in this process, the self-consistent requirement Eq.~(\ref{eq:sqrts}) is 
imposed at each step of the calculation of the eigenvalue. 

\begin{figure}[htb] 
\begin{center} 
\includegraphics[width=0.57\textwidth]{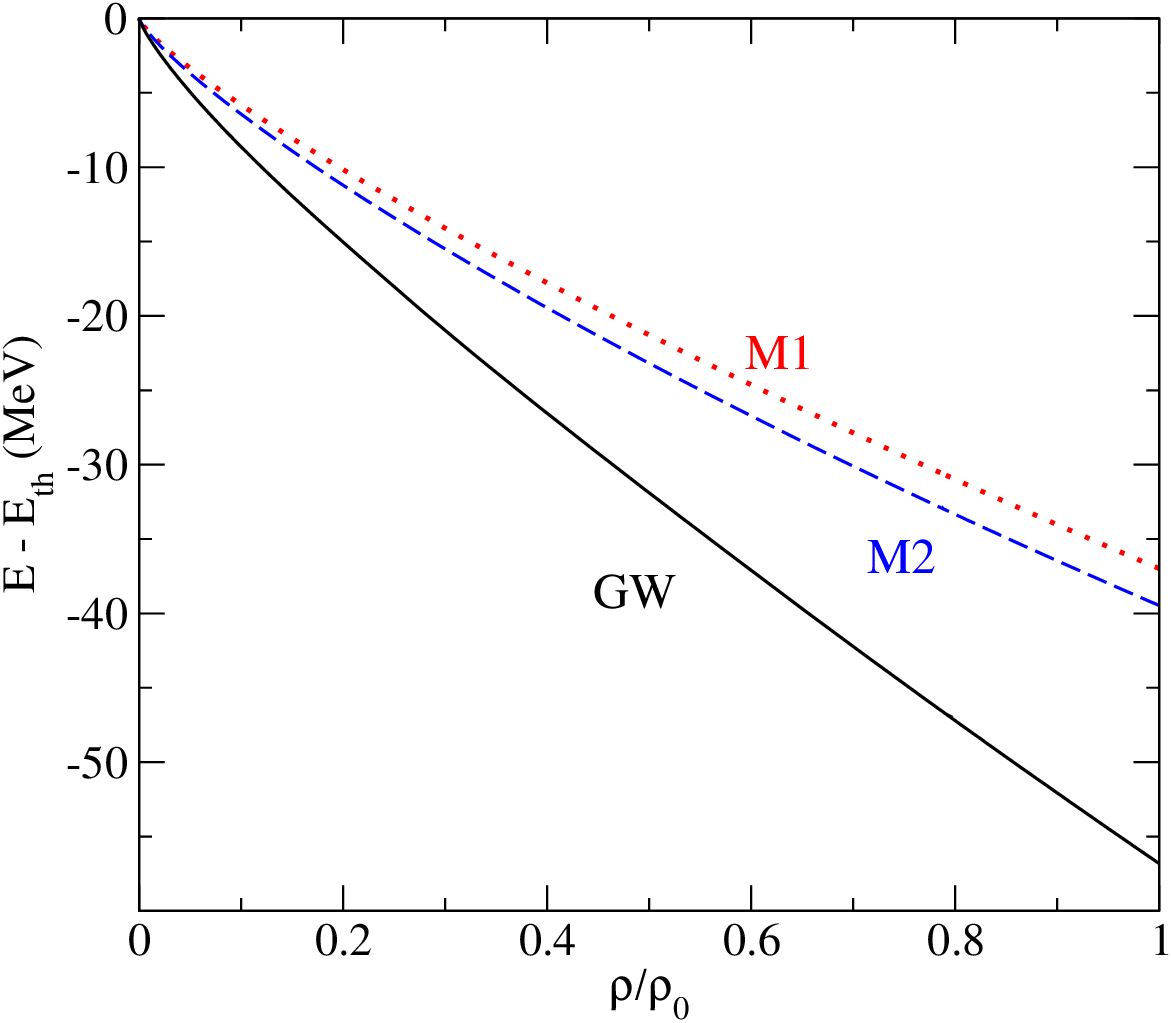} 
\caption{Subthreshold $\eta N$ energies probed by the $\eta$-nuclear potential 
as a function of the relative nuclear RMF density in Ca. Each of the three 
curves was calculated self-consistently according to Eq.~(\ref{eq:sqrts}) for 
a specific version of $\eta N$ subthreshold amplitude model, see text.}
\label{fig:srho} 
\end{center} 
\end{figure} 

In Fig.~\ref{fig:srho} we show the downward subthreshold energy shift 
$\delta\sqrt{s}\equiv E-E_{\rm th}$ as a function of the nuclear RMF density 
$\rho$ in Ca, calculated self-consistently according to Eq.~(\ref{eq:sqrts}) 
for $\eta N$ amplitude models M1, M2 and GW (see Table~\ref{tab:input}). 
The hierarchy of the three curves reflects the strength 
of the input Re\;$f_{\eta N}(\sqrt{s})$ in the subthreshold region, with 
threshold values listed in Table~\ref{tab:input}. It is clear that downward 
energy shifts of up to $\approx$55~MeV are involved in the present 
self-consistent calculations.

\section{Results and discussion} 
\label{sec:res} 

\begin{figure}[htb] 
\begin{center} 
\includegraphics[width=0.48\textwidth]{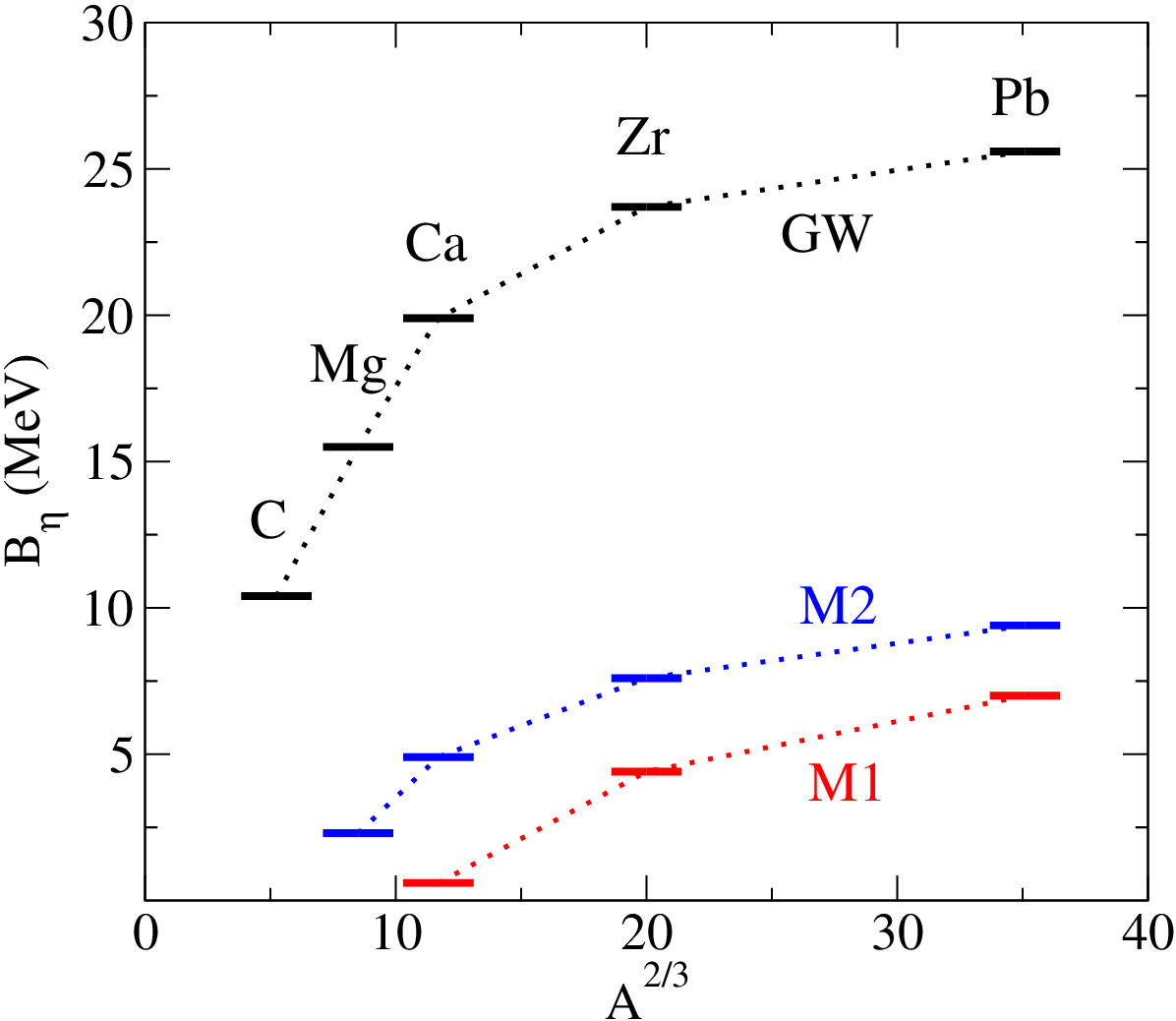} 
\includegraphics[width=0.48\textwidth]{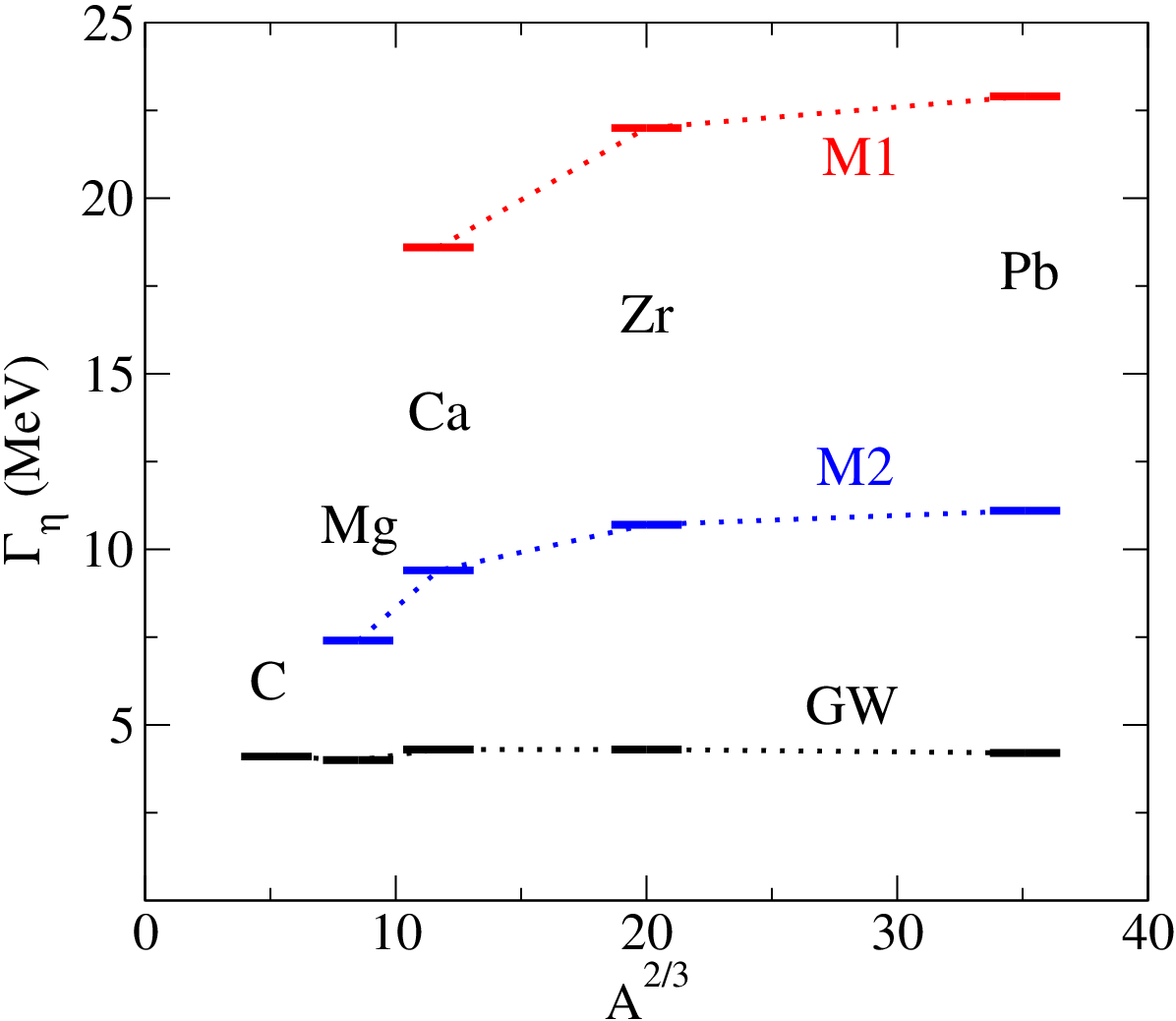} 
\caption{Binding energies (left) and widths (right) of $1s_{\eta}$ nuclear 
states across the periodic table calculated self-consistently using the 
M1, M2 and GW subthreshold $\eta N$ scattering amplitudes within a dynamical 
RMF scheme, see text.} 
\label{fig:bg-s} 
\end{center} 
\end{figure}

The methodology described in the last section was used to solve the KG 
equation (\ref{eq:KG}) for $\eta$-nuclear bound states across the periodic 
table. In this Letter we highlight the systematics of the $1s_{\eta}$ bound 
state and compare our treatment of subthreshold energy dependence with 
previous studies. A more detailed account plus extensions are given 
elsewhere \cite{CFGM13}. Three representative $\eta N$ amplitude models 
M1, M2 and GW (see Table~\ref{tab:input}) are employed here in order to span 
a wide range of $\eta N$ interaction strengths. Our main results are shown in 
Fig.~\ref{fig:bg-s} for binding energies $B_{\eta}$ and widths $\Gamma_{\eta}$ 
calculated for $1s_{\eta}$ nuclear states in core nuclei from  $^{12}$C to 
$^{208}$Pb. RMF equations of motion, along with the KG equation (\ref{eq:KG}), 
are solved self-consistently \cite{MFG06}, thereby allowing for core 
polarization by the $\eta$ meson (see Ref.~\cite{CFGGM11b} for the latest 
application to $K^-$ mesons). The core polarization effect on $B_{\eta}$ and 
$\Gamma_{\eta}$ was found in all cases displayed here to be less than 1~MeV. 
Therefore, the use of static nuclear densities is acceptable for not-too-light 
nuclear cores. 

Inspection of the l.h.s. of Fig.~\ref{fig:bg-s} reveals that for each of the 
three input $\eta N$ amplitude models the binding energy increases with $A$ 
and appears to saturate for large values of $A$. As in Fig.~\ref{fig:srho}, 
here too the hierarchy of the three curves reflects the strength of the 
Re\;$f_{\eta N}(\sqrt{s})$ input in the subthreshold region, with threshold 
values listed in Table~\ref{tab:input}. The M1 and M2 amplitudes are too weak 
to produce a $1s_{\eta}$ bound state in $^{12}$C, with the onset of binding 
for the weaker M1 amplitude deferred to around $^{40}$Ca. Of our three 
representative amplitudes, M1 is the closest one on shell to the Haider-Liu 
standard amplitude \cite{HL86,LH86,HL02} which was used by these authors 
to argue for $^{12}$C as the approximate onset of $\eta$-nuclear binding. 
In contrast, Re\;$f_{\eta N}(\sqrt{s})$ of the GW model is sufficiently strong 
to bind the $1s_{\eta}$ state in $^{12}$C and in lighter core nuclei, in spite 
of the suppression it undergoes here by forming its in-medium version and 
dealing with its energy dependence. The GW amplitude model even admits a 
$1s_{\eta}$ bound state in $^4$He with as low a binding energy as 1.2~MeV 
and a width of 2.3~MeV, both calculated using a static $^4$He density. 

Inspection of the r.h.s. of Fig.~\ref{fig:bg-s} reveals a trend for the three 
curves of calculated widths which is opposite to that observed on the l.h.s. 
for the calculated binding energies. Here, the GW model produces 
relatively small widths of order 4 MeV uniformly across the periodic table, 
whereas M1 and M2 give larger widths, particularly M1 with widths of order 
20~MeV. This reflects partly the energy dependence of Im\;$f_{\eta N}
(\sqrt{s})$ in the subthreshold region, which is quite distinct in each one 
of the three amplitude models, and partly the difference in the in-medium 
renormalization arising from the Re\;$f_{\eta N}(\sqrt{s})$ input. The latter 
point is readily understood by noting in Fig.~\ref{fig:srho} that the largest 
values of subthreshold downward energy shift are due to the GW subthreshold 
amplitude. This causes a particularly large reduction in the strength of the 
Im\;$f_{\eta N}(\sqrt{s})$ input for the GW amplitude model. 

\begin{table}[hbt] 
\begin{center} 
\caption{Static calculations of $1s_{\eta}$ binding energy ($B$) and width 
($\Gamma$) in $^{25}$Mg, using three $\eta N$ amplitude models (M1, M2, GW) 
with (YES) and without (NO) medium corrections from Eq.~(\ref{eq:WRW}), 
for several procedures of treating the energy dependence of $f_{\eta N}$. 
Energies and widths are given in MeV.} 
\begin{tabular}{llcccccc}
%\hline 
\noalign{\smallskip}\hline\noalign{\smallskip} 
Eq.~(\ref{eq:WRW}) & subthreshold & \multicolumn{2}{c}{M1} & 
\multicolumn{2}{c}{M2} & \multicolumn{2}{c}{GW} \\ 
\noalign{\smallskip}\cline{3-8}\noalign{\smallskip}  
& & $B_{\eta}$ & $\Gamma_{\eta}$ & $B_{\eta}$ & $\Gamma_{\eta}$ & $B_{\eta}$ & 
$\Gamma_{\eta}$  \\ 
%\hline 
\noalign{\smallskip}\hline\noalign{\smallskip} 
NO & $\delta\sqrt{s}=0$ & 3.2 & 37.4 & 17.3 & 37.8 & 81.8 & 62.7 \\ 
NO & $\delta\sqrt{s}=-30$ &--&--& 3.0 & 11.2 & 31.2 & 10.0 \\ 
NO & $\delta\sqrt{s}$ Eq.~(\ref{eq:sqrts}) &--&--& 3.2 & 10.6 & 23.8 & 7.4 \\ 
%YES & $\delta\sqrt{s}=-30$ &--&--& 1.9 & 8.2 & 16.7 & 4.2 \\ 
%NO & $\delta\sqrt{s}=-B_{\eta}$ & 3.2 & 34.5 & 10.9 & 21.4 & 30.8 & 9.8 \\ 
    &                    &     &      &      &      &      &      \\ 
YES & $\delta\sqrt{s}=0$ & 4.3 & 23.8 & 11.6 & 18.9 & 33.3 & 14.0 \\ 
YES & $\delta\sqrt{s}=-B_{\eta}$ & 3.8 & 21.7 & 8.3 & 13.2 & 19.4 & 5.8 \\ 
YES & $\delta\sqrt{s}$ Eq.~(\ref{eq:sqrts}) &--&--& 2.5 & 7.4 & 14.8 & 3.9 \\ 
\hline 
\end{tabular} 
\label{tab:Mg} 
\end{center} 
\end{table} 

Focusing on a given core nucleus, we show in Table~\ref{tab:Mg} results 
of static-density calculations in models M1, M2 and GW of the $1s_{\eta}$ 
state in $^{25}$Mg with (YES) and without (NO) employing the in-medium 
modification of Eq.~(\ref{eq:WRW}). The first row in each of the YES 
and NO groups lists results of using threshold amplitudes: 
$F_{\eta N}(\sqrt{s_{\rm th}},\rho)$ for YES and 
${\tilde f}_{\eta N}(\sqrt{s_{\rm th}})$ for NO. 
The self-consistency requirement imposed by Eq.~(\ref{eq:sqrts}) 
is used and comparison is made within each group with another procedure 
applied in previous studies to incorporate energy dependence. These are (i) 
a fixed 30~MeV downward shift applied to the free-space $\eta N$ amplitude 
$f_{\eta N}(\sqrt{s})$ by Haider and Liu (HL) \cite{HL02}; and (ii) shifting 
down self-consistently the energy argument of the in-medium $\eta N$ amplitude 
$F_{\eta N}(\sqrt{s},\rho)$ by the resultant $B_{\eta}$, as implemented for 
example by Garc\'{i}a-Recio et al. (GR) \cite{GR02}. 

The HL procedure is compared with ours in the second and third rows of the 
first (NO) group, using free-space amplitudes. Both do not produce binding 
for the weakest amplitude M1 and practically agree for M2, while disagreeing 
significantly for the strongest GW amplitude. By comparing each of these rows 
with the first row where free-space {\it threshold} amplitudes are used, it is 
seen the effects of accounting for energy dependence are substantial in both. 
The GR procedure is compared with ours in the second and third rows of the 
second (YES) group, using {\it in-medium} amplitudes. The GR procedure is 
found to give higher binding energies and widths than ours for all amplitude 
models tested here, particularly for the weaker M amplitudes where it is the 
only one that produces a $1s_{\eta}$ bound state for M1. The overall effects 
in this group of accounting for energy dependence with respect to using 
in-medium threshold amplitudes (first row of the YES group), however, are less 
substantial than in the preceding group, particularly for the M amplitudes. 

Of the three models used by us with in-medium amplitudes in 
Table~\ref{tab:Mg} (last line), only GW provides $B_{\eta}$ which is 
comparable with 
\begin{equation} 
B^{\rm exp}({_{~\eta}^{25}{\rm Mg}})=13.1\pm 1.6~{\rm MeV}, \;\;\;\;\; 
\Gamma^{\rm exp}({_{~\eta}^{25}{\rm Mg}})=10.2\pm 3.0~{\rm MeV}, 
\label{eq:Mg} 
\end{equation} 
deduced from the following $_{~\eta}^{25}$Mg interpretation of a peak reported 
by the COSY-GEM Collaboration \cite{Bud09}: 
\begin{eqnarray} 
p~+~{^{27}{\rm Al}} & ~\rightarrow & {_{~\eta}^{25}{\rm Mg}}~+~{^{3}{\rm He}} 
\nonumber \\ 
 &  &  ~\hookrightarrow ~ (\pi^{-}~+~p)~+~X,  
\label{eqnarray:Mg} 
\end{eqnarray}
with a decay induced by $\eta + n \to \pi^{-} + p$. Hence, if this peak 
assignment to a $1s_{\eta}$ state is correct,{\footnote{This has been 
contested recently by Haider and Liu who offered a different interpretation 
of the reported peak \cite{HL10}.}} then the underlying threshold value 
Re\;$a_{\eta N}$ must be rather large, close to 1 fm. Other procedures listed 
in Table~\ref{tab:Mg} for treating the subthreshold $\eta N$ energy dependence 
require considerably smaller values of Re\;$a_{\eta N}$. Finally, the 
relatively small value of width $\Gamma$ produced in the GW model should not 
be viewed as too restrictive since the total width must be larger than given 
in these models, owing to true $\eta NN$ absorption and two-pion production 
$\eta N\to\pi\pi N$ processes that are not accounted for by the models 
considered in the present work.

\section{Conclusions} 
\label{sec:sum}  

In this work we have demonstrated the importance of, as well as the subtleties 
involved in constructing self-consistent $\eta$-nucleus optical potentials 
that incorporate the strong subthreshold energy dependence of the underlying 
$\eta N$ scattering amplitude. Of the three $\eta N$ amplitude models studied 
here self-consistently, even the relatively weak attraction in model M1 with 
a threshold value Re\;$a_{\eta N}\approx 0.2$~fm requires going down to about 
35~MeV below threshold, as shown in Fig.~\ref{fig:srho}, in order to calculate 
reliably the $\eta$-nuclear optical potential $V_{\eta}(\rho)$ at central 
nuclear densities. This downward energy shift exceeds by far the downward 
shifts $-B_{\eta}$, with $B_{\eta}\lesssim 20$~MeV encountered in the 
self-consistent calculations of Garc\'{i}a-Recio et al. \cite{GR02}. 
The relatively large downward energy shifts in the present approach 
together with the rapid decrease of the free-space and in-medium $\eta N$ 
amplitudes lead to smaller than ever binding energies and widths with 
respect to those calculated in comparable models \cite{HL02,GR02,jido}. 
Thus, $_{~\eta}^{12}$C bound states are unlikely in models with threshold 
values Re\;$a_{\eta N}\lesssim 0.5$~fm, and as large a value as 
Re\;$a_{\eta N}\approx 0.9$~fm is required to reproduce the $_{~\eta}^{25}$Mg 
bound-state hint from the recent COSY-GEM experiment \cite{Bud09}. 
Complementarily, for as sufficiently large values of Re\;$a_{\eta N}$ as 
provided by the GW amplitude model, the calculated widths come out smaller 
than in other calculations. 

A value of Re\;$a_{\eta N}\sim 0.9$~fm is likely to yield a near-threshold 
$_{\eta}^{4}$He bound state, as found here using the GW amplitude model, 
but it is short of binding $_{\eta}^{3}$He. Stretching the limits of optical 
potential usage down to these light systems is of course questionable, and 
corresponding few-body calculations are highly needed to resolve such issues. 
Nevertheless, if one applies our subthreshold self-consistency scheme to 
$\eta$--$^{3}$He low-energy scattering, then a very large imaginary part 
that might indicate a nearby virtual state is found for the 
$\eta$--$^{3}$He cm scattering amplitude when using the GW amplitude model. 
This large imaginary part might be associated with the strong final-state 
interaction effects observed for the $\eta$--$^{3}$He system \cite{LNS88}. 
In contrast to previous estimates that assigned a value of Re\;$a_{\eta N}
\approx 0.5$~fm to describe such occurrence \cite{Willis97}, 
in our self-consistent calculations it requires substantially larger values, 
more likely around 0.9~fm.

\section*{Acknowledgements} 
We thank Maxim Mai for providing us with numerical tables of $\eta N$ 
scattering amplitudes from Ref.~\cite{MBM12}. A.G. thanks Colin Wilkin for 
useful and stimulating correspondence. J.M. acknowledges financial support 
within the agreement on scientific collaboration between the Academy of 
Sciences of the Czech Republic and the Israel Academy of Sciences and 
Humanities. This work was supported by the GACR Grant No. 203/12/2126, 
as well as by the EU initiative FP7, HadronPhysics3, under the SPHERE and 
LEANNIS cooperation programs.


\begin{thebibliography}{99} 

\bibitem{Gal13} For a recent review at HYP2012, see A.~Gal, Nucl. Phys. A 
(2013) http://dx.doi.org/10.1016/j.nuclphysa.2013.01.011 (arXiv:1301.2145). 

\bibitem{HL86} Q.~Haider, L.C.~Liu, Phys. Lett. B 172 (1986) 257. 

\bibitem{LH86} L.C.~Liu, Q.~Haider, Phys. Rev. C 34 (1986) 1845. 

\bibitem{HL02} Q.~Haider, L.C.~Liu, Phys. Rev. C 66 (2002) 045208. 

\bibitem{weise95} N.~Kaiser, P.B.~Siegel, W.~Weise, Phys. Lett. B 362 
(1995) 23; N.~Kaiser, T.~Waas, W.~Weise, Nucl. Phys. A 612 (1997) 297; 
T.~Waas, W.~Weise, Nucl. Phys. A 625 (1997) 287. 

\bibitem{IOV02} T.~Inoue, E.~Oset, M.J.~Vicente Vacas, Phys. Rev. C 65 (2002) 
035204. 

\bibitem{MBM12} M.~Mai, P.C.~Bruns, U.-G.~Mei{\ss}ner, Phys. Rev. D 86 (2012) 
094033. 

\bibitem{GW05} A.M.~Green, S.~Wycech, Phys. Rev. C 71 (2005) 014001, 
and references listed therein to earlier work by these authors. 

\bibitem{Arndt05} R.A.~Arndt, et al., Phys. Rev. C 72 (2005) 045202; 
see Table I for $\eta N$ scattering-length overview. 

\bibitem{SLM13} V.~Shklyar, H.~Lenske, U.~Mosel, Phys. Rev. C 87 (2013) 
015201. 

%\bibitem{LS06} M.F.M. Lutz, M. Soyeur, Nucl. Phys. A 773 (2006) 239, 
%and references listed therein. 

%\bibitem{Machner10} H.~Machner, Acta Phys. Polon B 41 (2010) 2221; see also 
%Refs.~\cite{MAMI12,COSY13} for more recent searches for $\eta$-$^3$He and 
%$\eta$-$^4$He bound states, respectively. 

\bibitem{LNS88} For $\eta$$^3$He: J.~Berger, et al. (LNS-SPES4), Phys. Rev. 
Lett. 61 (1988) 919; B.~Mayer, et al. (LNS-SPES2), Phys. Rev. C 53 (1996) 
2068; T.~Mersmann, et al. (COSY-ANKE), Phys. Rev. Lett. 98 (2007) 242301; 
J.~Smyrski, et al. (COSY-11), Phys. Lett. B 649 (2007) 258; C.~Wilkin, 
et al., Phys. Lett. B 654 (2007) 92. 

\bibitem{LNS94} For $\eta$$^4$He: R. Frascaria, et al. (LNS-SPES4), Phys. Rev. 
C 50 (1994) R537; N.~Willis, et al. (LNS-SPES3), Phys. Lett. B 406 (1997) 14; 
A.~Wro\'{n}ska, et al. (COSY-ANKE), Eur. Phys. J. A 26 (2005) 421; 
A.~Budzanowski, et al. (COSY-GEM Collab.), Nucl. Phys. A 821 (2009) 193.  

\bibitem{BTh06} S.D.~Bass, A.W.~Thomas, Phys. Lett. B 634 (2006) 368. 

\bibitem{MAMI12} F.~Pheron, et al. (MAMI), Phys. Lett. B 709 (2012) 21. 

\bibitem{COSY13} P.~Adlarson, et al. (WASA@COSY Collab.), Phys. Rev. C 87 
(2013) 035204. 

\bibitem{GR02} C.~Garc\'{i}a-Recio, T.~Inoue, J.~Nieves, E.~Oset, Phys. Lett. 
B 550 (2002) 47. 

\bibitem{jido} D.~Jido, H.~Nagahiro, S.~Hirenzaki, Phys. Rev. C 66 (2002) 
045202; H.~Nagahiro, D.~Jido, S.~Hirenzaki, Phys. Rev. C 68 (2003) 035205; 
D.~Jido, E.E.~Kolomeitsev, H.~Nagahiro, S.~Hirenzaki, Nucl. Phys. A 811 (2008) 
158. 

\bibitem{CFGGM11a} A.~Ciepl\'{y}, E.~Friedman, A.~Gal, D.~Gazda, J.~Mare\v{s}, 
Phys. Lett. B 702 (2011) 402. 

\bibitem{CFGGM11b} A.~Ciepl\'{y}, E.~Friedman, A.~Gal, D.~Gazda, J.~Mare\v{s}, 
Phys. Rev. C 84 (2011) 045206. 

\bibitem{FG12} E.~Friedman, A.~Gal, Nucl. Phys. A 881 (2012) 150. 

\bibitem{GM12} D.~Gazda, J.~Mare\v{s}, Nucl. Phys. A 881 (2012) 159. 

\bibitem{BGL12} N.~Barnea, A.~Gal, E.Z.~Liverts, Phys. Lett. B 712 (2012) 132. 

\bibitem{FG13} E.~Friedman, A.~Gal, Nucl. Phys. A 899 (2013) 60. 

\bibitem{Bud09} A.~Budzanowski, et al. (COSY-GEM Collab.), Phys. Rev. C 
79 (2009) 012201(R).

\bibitem{WRW97} T.~Waas, M.~Rho, W.~Weise, Nucl. Phys. A 617 (1997) 449. 

\bibitem{CFGM13} A.~Ciepl\'{y}, E.~Friedman, A.~Gal, J.~Mare\v{s}, 
in preparation. 

\bibitem{MFG06} J.~Mare\v{s}, E.~Friedman, A.~Gal, Nucl. Phys. A 770 
(2006) 84. 

\bibitem{HL10} Q.~Haider, L.C.~Liu, J. Phys. G: Nucl. Part. Phys. 37 (2010) 
125104. 

\bibitem{Willis97} N.~Willis, et al. (LNS-SPES3), Phys. Lett. B 406 (1997) 14. 

\end{thebibliography}
\end{document}